\documentclass[11pt]{article}
\newcommand{\BEQ}{\begin{equation}}
\newcommand{\EEQ}{\end{equation}}

\newcommand{\DEG}[1]{\mbox{$ #1^{\rm o}$}}
\newcommand{\eq}[1]{eq.~(\ref{#1})}
\newcommand{\gappr}{\hspace{0.3em}\raisebox{-0.6ex}{$\stackrel{>}{\sim}$}\hspace{0.3em}} 
\newcommand{\ETC}{\mbox{\em etc.\/ }}
\newcommand{\VIZ}{\mbox{\em viz.\/ }}
\newcommand{\CF}{\mbox{\em cf.\/ }}

\newcommand{\ETAL}{\mbox{\em et. al.\/ }}
\newcommand{\EG}{\mbox{\em e.g.\/ }}
\begin{document}
\title{\hfill {\small JHU-TIPAC 98013}\\[1mm]
Strongly Interacting Neutrinos and the Highest Energy Cosmic Rays}
\author{G. Domokos and S. Kovesi-Domokos\\[1mm]
Department of Physics and Astronomy\\
The Johns  Hopkins University\\
Baltimore, MD 21218\thanks{E-mail: skd@haar.pha.jhu.edu}}
\date{December 1998}
\maketitle
\begin{abstract}
Cosmic rays of energies larger than the Greisen--Zatsepin--Kuzmin (GZK)
cutoff may be neutrinos if they acquire strong interactions due to
a ``precocious unification'' of forces. A scenario for this to happen 
is outlined. There is no contradiction with precision measurements
carried out at LEP and SLAC. Observable consequences at LHC and
future neutrino detectors are discussed.
\end{abstract}
\vspace{10mm}
\begin{flushleft}
\end{flushleft} 
A substantial number of cosmic ray events has been detected in which the 
primary energy appears to exceed the Greisen-Zatsepin-Kuzmin (GZK)
cutoff, see 
\EG  \cite{takeda,bird,brooke,efimov} and Szabelski's recent 
review~\cite{szabelski}. The onset of the GZK cutoff itself
is somewhat uncertain: the energy of the rapid turnover of the spectrum 
depends on
several   details  (for instance, on the injection spectrum, \ETC),
see ref.~\protect\cite{hillschramm} for a modern treatment.
Nevertheless, it is unlikely that all the events reported can
be explained by fluctuations in the shower development.

A number of explanations of the ``anomalous'' events has been offered;
for an overview, \CF  the proceedings of the 1997 University
of Maryland workshop \protect\cite{giant}. In a recent
article, however,   Burdman,  Halzen and Gandhi
ref.~\protect\cite{burdman}  conclude that none of the 
explanations offered in the literature is a convincing one.

Given this situation, we reexamine a 
proposal put forward some time ago, \protect\cite{oldone1, oldone2}.
In those papers, we proposed that at sufficiently high energies, neutrinos 
(in general, leptons)
acquire some unspecified strong interaction and cause (possibly)  post-GZK 
showers. Since neutrinos are neutral, have small magnetic
moments and at low energies they have nothing but  weak interactions,
they can freely propagate through the \DEG{2.7}K background.
Consequently, they can reach us from cosmological distances.

There were  two weaknesses inherent in this suggestion. 
First, no mechanism has been offered
as to how the neutrinos get their strong interaction. Second, due to the
fact that we used a sharp ($\Theta$ function) threshold to turn on the
strong interaction, the resulting amplitude violated 
unitarity, \CF \protect\cite{burdman}.

The scenario we present  here is based on recent work designed to 
overcome the hierarchy problem. The authors of 
refs.~\protect\cite{Dimo1, Dimo2, Dimo3, Duda1, Duda2, Duda3} conjecture that
unification may take place at a much lower energy than originally 
suspected, perhaps at a few TeV ({\em ``precocious unification''\/}). Even 
though  there are  some problems with this approach (\EG
the long lifetime of the proton is still lacking a convincing
explanation), it offers some interesting possibilities regarding the
nature of the physics beyond the Standard Model.

The high unification masses obtained within the Standard Model and its
minimal supersymmetric extensions, 
see~\protect\cite{kim, amaldi, ellis, lang}, is 
the result  of the assumption of a ``desert'' between -- approximately -- the
weak scale and the GUT scale. If one wants to accomplish
unification at a lower energy, one has to postulate  the
existence of a rapidly increasing density of states somewhere 
above the weak scale. 

For the purpose of the present paper, it is irrelevant whether the
rapid increase of the density of states is due to the existence of
extra dimensions, due to ``stringy effects'' becoming relevant
at lower energies or to something else.

Let us now assume that we have a rapidly increasing level density, which has
approximately the same form as in string theories. We choose:
\BEQ
d\left(m_{n}\right) \sim \left( \frac{m_{n}^{2}}{m_{0}^{2}}\right)^{-\alpha}
\exp \left( \frac{m_{n}^{2}}{m_0^{2}}\right)^{\rho},
\label{density}
\EEQ
see for instance \protect\cite{Green}. Here, $m_{n}$
stands for the mass of the resonance at the $n^{th}$ level of a
string model, $m_{0}$ is the characteristic energy scale of the 
model. A  numerical prefactor has been omitted: it is model dependent
and it it does not play a significant role in what follows. Similarly,
the exponent  $\alpha$ is model dependent. We found, however that varying
the exponent does not affect the results significantly;
for the sake of definiteness, we settled with $\alpha = 2$.

We quoted an expression corresponding to the asymptotic form of 
the level density in a string model. Due to the fact that  at this stage
we want to understand the qualitative behavior of the neutrino
cross section in the scenario just outlined, the present expression should be
adequate. In most string models, the value of the exponent $\rho$
is 1/2. Nevertheless, we wanted to explore more rapidly rising 
level densities as well; all the estimates were carried out with
$\rho =1/2$ and $\rho = 1$.

In order to estimate the neutrino--quark cross section, we take into account
the contribution of the resonances to the imaginary part of the
forward neutrino--quark amplitude. The invariant Breit-Wigner formula
is used for a single resonance. In this way, we get the contribution of 
a single resonance of mass squared $s_{n}$ at level $n$ of a string model:
\BEQ
Im B_{n} = \frac{1}{\pi} 
\frac{s \Gamma_{n} s_{n}^{1/2}}{\left(\hat{s}-s_{n}\right)^{2}+\Gamma_{n}^{2}s_{n}}
\label{breitwigner}
\EEQ
In this equation $\hat{s}$ stands for the CM energy squared of the
neutrino-quark system.

The total cross section due to the ``new physics'' is then obtained
by multiplying \eq{breitwigner} with the level density and summing
over the levels. We assume that $\Gamma_{n}s_{n}^{1/2}$ grows linearly with
$s_{n}$. \VIZ
$\Gamma_{n}s_{n}^{1/2} = \gamma_{0}s_{n}$ and that the resonances lie on a
linear Regge trajectory, $s_{n} = s_{0} (n - 1)$. Further,
we identify $s_{0}= m_{0}^{2}$ in \eq{density}. (In a string model, both
quantities are related to the string tension; there may be prefactors of
order unity which we ignore here.) The linear growth of the width
can be understood in intuitive terms by noticing that a resonance
preferentially decays  into channels with the largest phase space
available. The number of such channels is roughly proportional to
the mass of the resonance, hence $\Gamma_{n}s_{n}^{1/2} \propto s_{n}$.
(It is known, for instance that the total widths of baryon
resonances grow approximately linearly with their masses 
\CF \cite{schonberg}.)

We now notice that $\gamma_{0}$ is likely to be rather small,
\CF~\cite{Dimo1}.
 Hence,  the quantity $ Im B_{n}$ may be replaced by a $\delta$-function
and the summation over the levels by an integration.
In this approximation  the neutrino--quark
cross section becomes:
\BEQ
\hat{\sigma} \approx \frac{1}{s_{0}}d\left(\hat{s}\right)
\label{quarksigma}
\EEQ
Finally, one has to integrate \eq{quarksigma} over the momentum
distribution of quarks within the target nucleon, using the
relation $\hat{s}= xs$. Little or nothing is known about the evolution 
of structure functions into a region of rapidly increasing level
density. In order to get some idea, we chose a ``generic''
structure function, of the form:
\BEQ
S(x) = A x^{-\gamma}\left( 1-x\right)^{\delta}
\label{structurefunction}
\EEQ
Inspired by the Duke--Owens parametrization of the structure functions,
\CF~\cite{bargerphillips} we chose $ A=2, \gamma = 0.6, \delta =3.5$.
None of the results turned out to be very sensitive 
to the precise choice of these parameters. The final expression
of the neutrino-nucleon cross section is given by:
\BEQ
\sigma = \int_{x_{0}}^{1}dx S(x)\hat{\sigma}(xs)
\label{sigmatot}
\EEQ
The infrared cutoff was chosen as $x_{0}= s_{0}/s$; in this way,
the ``new physics'' begins to manifest itself for $s\gappr s_{0}$.
In order to test the scheme developed, let us
assume that we want the cross section to grow to 
$\sigma \approx 1/\Lambda_{QCD}^{2}$ around the GZK cutoff,
say, $s= 2\times 10^{4}{\rm TeV}^{2}$. (This corresponds to a
laboratory energy, $E_{L}\approx 10^{19}$eV.) On taking 
$\Lambda_{QCD}=200$MeV, one gets $s_{0}\approx 3$TeV for
$\rho =1/2$ in \eq{density} and $s_{0} \approx 30$TeV
for $\rho =1$, respectively.
In the following figures we display the cross sections resulting from
\eq{sigmatot} taking  $\rho=1/2$ and $\rho = 1$ in \eq{density}.
%
\input epsf.tex
\noindent
\begin{figure}
\vspace{15mm}
\begin{center}
\epsfxsize=100mm \epsfbox{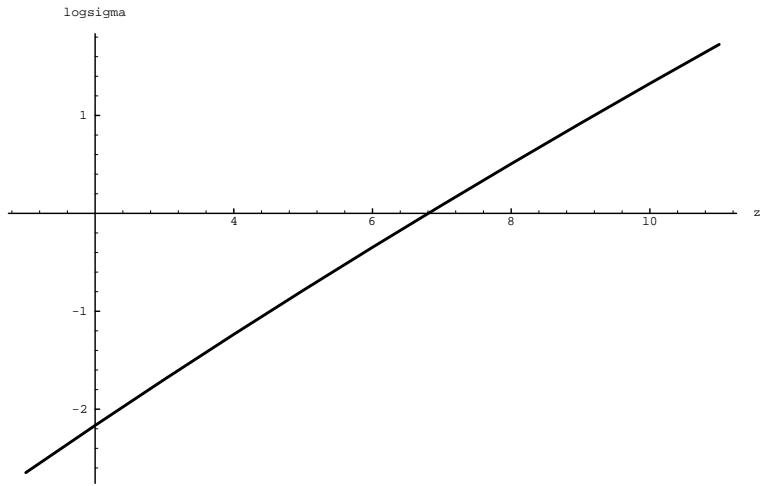} \vspace{-32mm}
\caption{Decimal logarithm of the  $\nu-N$ cross section measured in 
millibarns; $\rho = 1/2$, $s_{0}^{1/2} = 3$TeV; $z=s/s_{0}$}
\end{center} \label{fig:smallrho}
\end{figure}
%
\begin{figure} 
\vspace{15mm}
\begin{center}
\epsfxsize=100mm \epsfbox{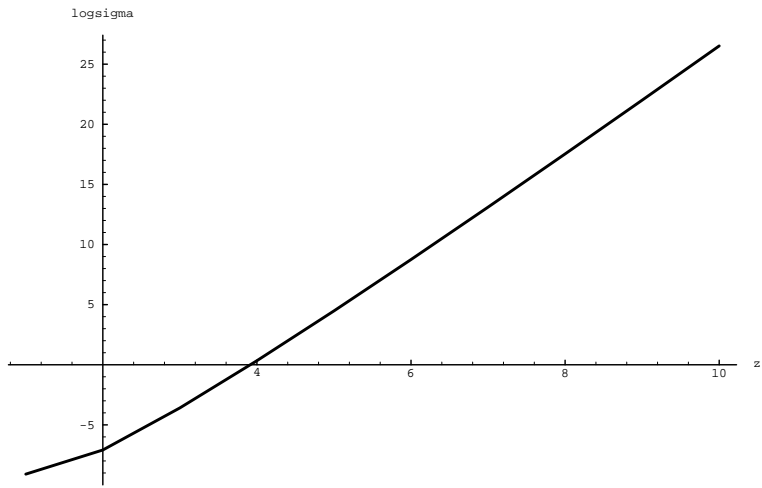}
\vspace{-32mm}
\caption{Same as Fig.~1, but  with $\rho=1$, $s_{0}^{1/2}=30$TeV}
\end{center}\label{fig:bigrho}
\end{figure}
%
%

One sees that, for all practical
purposes, the neutrino-nucleon cross section is dominated by the 
 exponential growth of the level density; hence, the choice of the 
exponents in \eq{density} and \eq{structurefunction} is not critical.

The cross sections rise very rapidly and, as it is always the case
with tree amplitudes containing high spin particles in intermediate
states, it will violate unitarity at a certain energy. The value of that
energy is, however, difficult to determine at this stage. This is due to
the fact that, like in any string model, at level $n$, resonances 
with spins $0 \leq s \leq n$ are exchanged. (In some  models the lower
limit may be $1$ and the upper limit $n\pm 1$, but this fact does
not affect the general situation.) As a consequence, the usual
semiclassical estimates of the unitarity limits of cross sections
are not directly applicable. Only by calculating loop corrections
will one be able to estimate the unitarity limits. 

It is to be
emphasized that {\em individual partial waves} need not violate the
unitarity bounds. In order to illustrate this point, consider a
simple Veneziano amplitude,
\BEQ
A(s,t) = {\cal B} \left( 1-\alpha(s), 1-\alpha(t) \right),
\label{veneziano}
\EEQ
where $ {\cal B} (x,y)$ is the Euler beta function and $\alpha$ is a linear
Regge trajectory.
This amplitude has a level density asymptotically described by \eq{density}
with $\rho = 1/2$. By using Stirling's formula, one  obtains an 
 asymptotic estimate for the partial wave amplitudes.
In particular, one finds for the $S$ wave amplitude:
\BEQ
A_{0}(s)\sim \left(2\pi \alpha(s)\right)^{-1/2}\quad 
\left( \alpha(s) \gg 1\right).
\label{partial}
\EEQ
Clearly, the bound given by 
$\left| \exp (i\delta_{0}) \sin \delta_{0}\right|\leq 1$ is satisfied
asymptotically. (In order to arrive  at the result given by \eq{partial}, 
one has
to avoid the poles of the $\Gamma$ function, \EG by sending $s$
to infinity along a ray in the complex plane, 
$s = |s|\exp i\phi, \quad \phi \not= 0$.)

{\em Are the precision results of the Standard Model affected\/?}
This is an important question: the effects of the ``new physics''
are expected to be observable even below the characteristic energy
scale, \CF Goldberg and Weiler, ref.~\cite{goldbergweiler}.
As in discussing the unitary bounds before, we are unable to 
estimate loop effects at this stage. However, one can estimate 
final state interactions due to the ``new physics''. We use the
scattering length approximation to the $l^{th}$ partial wave, 
\BEQ
k^{(2l+1)} {\rm cot}\delta_{l}\approx \frac{1}{a_{l}}
\label{scatteringlength}
\EEQ
This should be reasonably accurate: typical CMS energies at precision
measurements carried out at LEP and SLAC are of the order of
$100$GeV, whereas the characteristic energy of the ``new physics''
is on the TeV scale. A reasonable estimate for the scattering
length is $1/a \approx \sqrt{s_{0}}$. The effect of the final
state interactions is given by the  formula,
\CF \cite{goldbergerwatson}:
\BEQ
w_{l} \approx w_{l}^{(0)} \left( 1 + \left( k^{(2l+1)}a_{l}\right)^{2}\right),
\label{finalstate}
\EEQ
In \eq{finalstate}, $w$ is a transition probability in a given partial wave
$l$, (for instance, $Z \rightarrow \nu \bar{\nu})$  and $w^{0}$ is the
same transition probability calculated ignoring the final state 
interaction. On using the estimate $\sqrt{s_{0}}\approx 3$TeV,
one finds that with $k \approx 45$GeV (half of the mass of the $Z$),
the $S$ wave final state ``enhancement factor'' differs from unity by
about $10^{-4}$ or so. This is to be compared with a typical error
of a precision measurement (for instance, the decay width of a
weak gauge boson) which is of the order of $0.1\%$,
\cite{pdg}. The effect is, 
of course, much smaller in higher partial waves or for
a higher characteristic energy, \CF \eq{finalstate}.

The scenario outlined here has some observable consequences.
They, in turn, depend on whether the characteristic energy
is of a few TeV or a few tens of TeV.
For a characteristic energy of a a few TeV,
\begin{itemize}
\item there should be measurable deviations from standard model predictions
in the decay rates of weak gauge bosons if the accuracy of the measurements
 can be increased by about an order of magnitude,
\item one expects spectacular phenomena of new particle production
and/or strong violation of Feynman scaling due to the rapid rise of the level
density at LHC. Due to the precocious unification, one expects a copious
production of leptons as well as hadrons.
\end{itemize}
For a high characteristic energy ($\simeq 30$TeV), this phenomenon 
may not take place at the LHC. However, new phenomena will be observed
at nonaccelerator experiments, such as neutrino telescopes and at 
orbiting detectors (OWL, Airwatch).
\begin{itemize}
\item It was pointed out that orbiting detectors should be sensitive to
neutrino interactions, see~\cite{orbiting}. If the cross section
of neutrino interactions shows a rapid rise, there should be a
corresponding rise of the impact parameter of the incident neutrino
with respect to the center of the Earth
at which showers generated by them can be observed. Accordingly,
there should be a cutoff in the spectrum of upward going neutrinos
observed in underground, under water or under ice detectors.
The neutrino induced showers develop 
rapidly within the Earth and they  degenerate by the time they
would reach the detector. By contrast, the number of showers of
grazing incidence  should increase.
\item This scenario resolves the ``energy crisis'' caused by
some schemes purporting to explain the highest energy cosmic rays,
such as~\cite{oldone1}. It is difficult enough to construct mechanisms
by means of which protons are accelerated to energies of the order of
$10^{20}$eV, \CF Norman \ETAL,~\cite{norman}. If one prefers 
the highest energy cosmic rays to be neutrinos originating from 
the decay of pions and kaons, one needs protons to be accelerated
to energies a few orders of magnitude even higher: at such energies, several
hundred light hadrons are produced and, on the average, the
available primary energy is shared equally between them. 
By contrast, in a scenario
involving precocious unification, once the CMS energy of the
accelerated protons in an active galactic nucleus or in a similar
site of intense proton acceleration reaches the characteristic energy,
neutrinos are produced at multiplicities comparable to hadrons.
Hence, there is no need to postulate proton energies several orders
of magnitude larger.
\end{itemize} \vspace{5truemm}
We thank L.~Madansky and C.~Norman for several conversations on the
highest energy cosmic rays.

\end{document}